\documentclass[9pt,twocolumn,twoside]{osajnl}
\usepackage{cuted}
\usepackage{color, soul}
\journal{ao}

\setboolean{shortarticle}{false}

\title{A flexible, fast and benchmarked vectorial model for focused laser beams}
\author[1,*]{Qingfeng Li}
\author[1]{Maxime Chambonneau}
\author[1]{Markus Blothe}
\author[1,2]{Herbert Gross}
\author[1,2]{Stefan Nolte}

\affil[1]{Institute of Applied Physics, Abbe Center of Photonics, Friedirich-Schiller-University Jena, Albert-Einstein-Str. 15, 07745 Jena, Germany}

\affil[2]{Fraunhofer Institute for Applied Optics and Precision Engineering, Albert-Einstein-Str. 7, 07745 Jena, Germany}

\affil[*]{Corresponding author: qingfeng.li@uni-jena.de} 

\begin{abstract}
	In-bulk processing of materials by laser radiation has largely evolved over the last decades and still opens up new scientific and industrial potentials. The development of any in-bulk processing application relies on the knowledge of laser propagation and especially the volumetric field distribution near the focus. Many commercial programs can simulate this, but, in order to adapt them, or to develop new methods, one usually needs to create a specific software. 
	Besides, most of the time people also need to measure the actual field distribution near the focus to evaluate their assumptions in the simulation. To easily get access to this knowledge, we present our high-precision field distribution measuring method and release our in-house software InFocus \cite{InFocus}, under the Creative Commons 4.0 License. Our measurements provide 300-nm longitudinal resolution and diffraction limited lateral resolution. The in-house software allows fast vectorial analysis of the focused volumetric field distribution in the bulk. The simulations of light propagation under different conditions (focusing optics, wavelength, spatial shape, propagation medium) are in excellent agreement with propagation imaging experiments. The aberrations provoked by the refractive index mismatch as well as those induced by the focusing optics are both taken into account. The results indicate that our proposed model is suitable for the precise evaluation of energy deposition.
\end{abstract}

\setboolean{displaycopyright}{true}

\begin{document}

\maketitle

\section{Introduction}\label{section:1}

In the last two decades, laser processing in the bulk of optical materials has attracted intensive attention in a wide range of academic researches and industrial engineering. In-volume laser direct writing enables precise three-dimensional structuring and has allowed many innovative applications that include the fabrication of channels \cite{bellouard2004fabrication, osellame2007integration, maselli2009femtosecond, he2010direct}, waveguides \cite{davis1996writing, chambonneau2016writing, pavlov2017femtosecond, gebremichael2020double,wang2020nanosecond}, gratings \cite{chambonneau2018inscribing}, data storage \cite{zhang2014seemingly}, and photonics quantum gates \cite{crespi2011integrated, lammers2019embedded}. The nature of the in-bulk processing also innovates new manufacturing procedures such as bonding \cite{richter2016laser,zhang2017interface, Cvecek2019, penilla2019ultrafast, Chambonneau2020} and dicing \cite{Meyer2019} of brittle materials. Among all potential applications, the precise description and control of the laser focusing and the energy deposition are crucial. Numerous experimental and theoretical investigations on the laser propagation and the energy absorption have been intensively carried out \cite{Couairon2007, berge2007ultrashort, gamaly2013generation,fedorov2016accessing, sahoo2020dynamic}. 

The propagation of the electromagnetic (EM) field can be rigorously described by the finite difference time domain (FDTD) method \cite{liu2000three}, however in general it requires significant computational resources. Considerable effort has been devoted to designing propagation equations that on one hand preserve their computational simplicity and on the other hand preserve the correct description of nonparaxial and vectorial effects. In the nonlinear propagation regime, only until recently, the unidirectional Hertz vector propagation equation (UHPE) \cite{couairon2015propagation, PhysRevE.100.033316} was derived to provide a seamless transition from Maxwell’s equations to the various envelope-based models, which considerably reduces the computational time. Simulations of the UHPE, require starting from input conditions, i.e., from the Hertz vector in a plane $z = z_0$. When the focusing elements have a high numerical aperture (NA), the input conditions are then determined by a phase correction to the field that simulates the action of the focusing element. The input conditions for the UHPE were constructed by a detailed calculation of diffraction by vectorial diffraction integrals (VDIs) \cite{varga2000focusing1,varga2000focusing2}, i.e. the linear propagation model. However, even though the vectorial effects have been considered by VDIs, the residual aberrations of the focusing elements are often ignored, which leads to a deviation from the correctness in the real laser processing conditions.

The purpose of this work is to accurately describe these input conditions by taking into account all the potential aberrations that may occur in the laser in-bulk focusing, and to allow for a fast analysis with a proper transformation of the VDIs. 

One of the most widely used integral for analyzing the vectorial diffraction is the Debye-Wolf integral. As demonstrated by Leutenegger \emph{et~al.} \cite{leutenegger2006fast}, the 3D vectorial field distribution at the focus can be computed plane by plane under a proper transformation of the original Debye-Wolf integral. At a given axial position, the EM field in this plane is obtained by a two-dimensional Fourier transform. Lin \emph{et~al.} \cite{Lin2012} have also demonstrated that this method is applicable to focusing through an interface between two media of mismatched refractive index. In this paper, we further adapt this method to the real lens conditions and provide a fast analysis tool for the evaluation of the actual EM field distribution at the focus of the lens whose residual aberrations cannot be neglected. Meanwhile, a non-destructive experimental method is introduced to provide 300-nm longitudinal and diffraction limited lateral resolution measurements of the in-bulk volumetric intensity distribution. Our numerical methods are benchmarked with experiments relying on propagation imaging under various conditions (focusing optics, wavelength, spatial shape, propagation medium).

\section{Model descriptions} \label{section:2}
\subsection{Three-dimensional Fourier Transform (3D-FT) representation of the field vectors near the focus}
Using the form developed by Richards and Wolf \cite{richards1959electromagnetic}, the time-dependent electric and magnetic fields ($\mathbf{E}$ and $\mathbf{H}$) in the image regime of a system can be expressed by \eqref{eq:1}. Here $\mathbf{e}$ and $\mathbf{h}$ are the time-independent electric and magnetic vectors, $\omega$ is the angular frequency.
\begin{equation}\label{eq:1}
	\begin{aligned}
		\mathbf{E}(x, y, z, t)&=\Re\{\mathbf{e}(x, y, z)e^{-i\omega t}\},\\
		\mathbf{H}(x, y, z, t)&=\Re\{\mathbf{h}(x, y, z)e^{-i\omega t}\}.
	\end{aligned}	
\end{equation}
At any point $\textbf{P}(x,y,z)$ in the image space, the electric and magnetic vectors $\mathbf{e}$ and $\mathbf{h}$ can be expressed in the form as a summation of the plane waves that are leaving the aperture:
\begin{equation}
	\begin{aligned}\label{eq:2}
		\textbf{e}(x, y, z)&=-\frac{ik}{2\pi}\iint_\Omega\frac{\textbf{a}(s_x, s_y)}{s_z}e^{ik[\textbf{$\Phi$}(s_x, s_y)+s_xx+s_yy+s_zz]}\,ds_x\,ds_y,\\
		\textbf{h}(x, y, z)&=-\frac{ik}{2\pi}\iint_\Omega\frac{\textbf{b}(s_x, s_y)}{s_z}e^{ik[\textbf{$\Phi$}(s_x, s_y)+s_xx+s_yy+s_zz]}\,ds_x\,ds_y
	\end{aligned}
\end{equation}
where $\mathbf{\Phi}(s_x, s_y)$ is the aberration function describing the optical path difference between the aberrated and the spherical wavefront along $\mathbf{s}$. Here $\mathbf{s}$ is a unit vector pointing from a point in the exit aperture to the focus, $\mathbf{a}$ and $\mathbf{b}$ are the electric and magnetic strength vectors of the unperturbed electric and magnetic fields in the exit aperture, $k$ is the wave number, and $\Omega$ is the solid angle formed by all the geometrical optical ray. The phase factor shown in \eqref{eq:2} contains two parts, one is the scalar product of vector $\mathbf{s}$ and vector $\mathbf{r}_p$, another is the vectorial aberration function. In this section henceforth we only discuss the electric field since, apart from the strength vector, the two equations in \eqref{eq:2} are identical. 

\begin{figure}
	\centering
	\includegraphics[width=\linewidth]{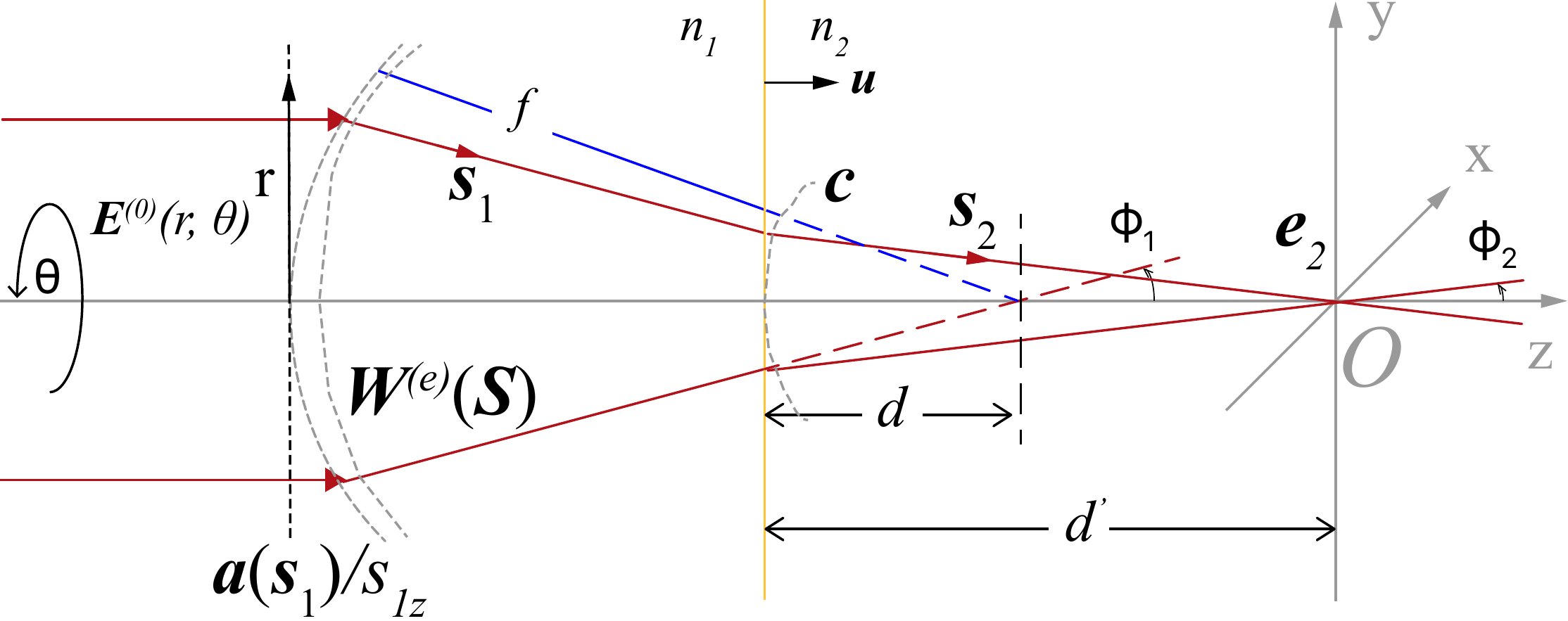}
	\caption{Diagram showing laser focused by a lens into two media separated by a planar interface.}\label{fig:1}
\end{figure}

Now let us consider a laser in-bulk focusing scenario. As shown in Fig.~\ref{fig:1}, after the focusing element, this configuration consists of materials 1 and 2 with refractive indices $n_1$ and $n_2$, respectively.

In material 1 and at the interface ($z=-d$), the electric field is given by
\begin{equation}\label{eq:3}
	\begin{aligned}
		\textbf{e}_1(x,y,-d)&=-\frac{ik_1}{2\pi}\iint_\Omega\textbf{C}(\vec{\mathbf{s}})\\
	&\times\exp[ik_1(s_{1x}x+s_{1y}y-s_{1z}d)]\,ds_{1x}\,ds_{1y}	
	\end{aligned}
\end{equation}
where 
\begin{equation}
	\textbf{C}(\vec{\mathbf{s}})=\frac{\textbf{a}(s_{1x},s_{1y})e^{ik_1\Phi(s_{1x},s_{1y})}}{s_{1z}}
\end{equation}\label{eq:4}
by combining the strength vector $\mathbf{a}(s_{1x},s_{1y})$ and the aberration phase factor $e^{ik\Phi(\vec{\mathbf{s}})}$ into the new complex strength vector $\textbf{C}(\vec{\mathbf{s}})$. The remaining phase factor in \eqref{eq:3} thus contains only the scalar product of vector \textbf{s} and vector \textbf{r}$_p$.

Since there is no optical coating for all the cubical materials presented in this paper, we assume that each plane wave components refraction at the interface obeys the Fresnel equations. To determine the transmitted field in the second material, we also assume that the field in the second material is constructed by the superposition of refracted plane waves. As the complex strength vector of the plane wave upon the interface is described as $\textbf{C}(\vec{\textbf{s}})$, the strength vector of the transmitted plane wave can be described as a linear function of $\textbf{C}(\vec{\textbf{s}})$, i.e, $\textbf{T}\cdot\textbf{C}(\vec{\textbf{s}})$, where $\mathbf{T}$ is a refraction operator which is a function of angle of incidence and $n_1$, $n_2$. Therefore, the transmitted field in the second material can be written as
\begin{equation}\label{eq:5}
	\begin{aligned}
		\textbf{e}_2(x,y,-d)&=-\frac{ik_1}{2\pi}\iint_{\Omega_1}\textbf{T}\cdot\textbf{C}(\vec{\textbf{S}})\\
		&\times\exp[ik_1(s_{1x}x+s_{1y}y-s_{1z}d)]\,ds_{1x}\,ds_{1y}		
	\end{aligned}
\end{equation}

On the other hand, as T\"or\"ok \emph{et~al.} \cite{torok1995electromagnetic} suggested, we can also represent the field in the second material again as superposition of plane waves, which is a solution of time-dependent wave equation and can be written as
\begin{equation}\label{eq:6}
	\textbf{e}_2(\textbf{r}_p)=-\frac{ik_2}{2\pi}\iint_{\Omega_2}\textbf{F}(\vec{\mathbf{s}_2})\exp(ik_2\vec{\textbf{s}}_2\cdot\mathbf{r}_p)\,ds_{2x}\,ds_{2y}.
\end{equation}
One can notice that \eqref{eq:5} is the boundary condition of \eqref{eq:6}. Let us now establish the relation between $\vec{\textbf{s}_1}$ and $\vec{\textbf{s}_2}$.

According to the law of refraction,
\begin{equation}\label{eq:7}
	k_1(\vec{\textbf{u}}\times\vec{\textbf{s}_1})=k_2(\vec{\textbf{u}}\times\vec{\textbf{s}_2}),
\end{equation}
where $\vec{\textbf{u}}$ is the unit vector that is normal to the interface. When a planar interface is presented $\vec{\textbf{u}}=(0,0,1)$, and we have
\begin{equation}\label{eq:8}
	k_1s_{1x}=k_2s_{2x},\,\,\,\,
	k_1s_{1y}=k_2s_{2y}.
\end{equation} 
By taking the coordinate transformation, \eqref{eq:6} yields
\begin{equation}\label{eq:9}
	\begin{aligned}
		\mathbf{e}_2(\mathbf{r}_p)&=-\frac{ik_2}{2\pi}\iint_{\Omega_1}\mathbf{F}(\vec{\mathbf{s}_2})\\
		&\times\exp(ik_2\vec{\mathbf{s}_2}\cdot\vec{\mathbf{r}_p})\mathbf{J}_0(s_{1x},s_{1y};s_{2x},s_{2y})\,ds_{1x}\,ds_{1y},		
	\end{aligned}
\end{equation}
where $\mathbf{J}_0$ is the Jacobian of the coordinate transformation obtained from \eqref{eq:8}:
\begin{equation}\label{eq:10}
	\mathbf{J}_0=\left(\frac{k_1}{k_2}\right)^2,
\end{equation}
As \eqref{eq:9} must satisfy the boundary condition represented by \eqref{eq:5}, we have
\begin{equation}\label{eq:11}
	\mathbf{F}(\vec{\mathbf{s}_1},\vec{\mathbf{s}_2})=(\frac{k_2}{k_1})\mathbf{T}\cdot\mathbf{C}(\vec{\mathbf{s}})\exp[id(k_2s_{2z}-k_1s_{1z})].
\end{equation}

By substituting \eqref{eq:11} into \eqref{eq:9} we obtain the electric field in the second material:
\begin{equation}
	\begin{aligned}\label{eq:12}
	\mathbf{e}_2(x,y,z)&=-\frac{ik^2_2}{2\pi k_1}\iint_{\Omega_1}\mathbf{T}\cdot\mathbf{C}(\vec{\mathbf{s}})\\
	&\times\exp[id(k_2s_{2z}-k_1s_{1z})]\exp(ik_2s_{2z}z)\\
	&\times\exp[ik_1(s_{1x}x+s_{1y}y)]\,ds_{1x}\,ds_{1y}.
	\end{aligned}
\end{equation}
The first phase factor $\exp[id(k_2s_{2z}-k_1s_{1z})]$ stands for the aberration induced by the interface. The second phase factor $\exp(ik_2s_{2z}z)$ accounts for the phase accumulation when propagating along the z-axis, and the third term $\exp[ik_1(s_{1x}x+s_{1y}y)]$ represents the phase difference of the wave front at off-axis points (x, y, z) with respect to the on-axis point (0, 0, z).
Depending on the chosen coordinates, the following forms of the wave vectors are equivalent:
\begin{equation}\label{eq:13}
	\begin{aligned}
		&\vec{k_1}=\begin{pmatrix}
			k_{1x}\\
			k_{1y}\\
			k_{1z}
		\end{pmatrix}=k_1	
		\begin{pmatrix}
			-s_{1x}\\
			-s_{1y}\\
			s_{1z}	
		\end{pmatrix}=k_1
		\begin{pmatrix}
			-\sin\phi_1\cos\theta\\
			-\sin\phi_1\sin\theta\\
			\cos\phi_1	
		\end{pmatrix},\,\,\\\\
	&\vec{k_2}=\begin{pmatrix}
		k_{2x}\\
		k_{2y}\\
		k_{2z}
	\end{pmatrix}=k_2	
		\begin{pmatrix}
			-s_{2x}\\
			-s_{2y}\\
			s_{2z}	
		\end{pmatrix}=k_2
		\begin{pmatrix}
			-\sin\phi_2\sin\theta\\
			-\sin\phi_2\cos\theta\\
			\cos\phi_2	
		\end{pmatrix}.		
	\end{aligned}
\end{equation}
Therefore, $\,ds_{1x}\,ds_{1y}$ can be written as $\,dk_{1x}\,dk_{1y}/k_1^2$ and the interface-induced aberration can be written as 
\begin{equation}\label{eq:14}
	\Psi(\phi_1,\phi_2,d)=d(n_2\cos\phi_2-n_1\cos\phi_1).
\end{equation}
The spherical-polar form of the complex strength vector after the interface is
\begin{equation}\label{eq:15}
	\mathbf{C}(\phi_1,\phi_2,r,\theta)=\mathbf{T}(\phi_1,\phi_2,\theta)\cdot\mathbf{a}(r,\theta)e^{ik_1\Phi(r,\theta)}/\cos\phi_1.
\end{equation}
By using $\mathbf{c}(\phi_1,\phi_2,r,\theta)$ to brief note $\mathbf{T}(\phi_1,\phi_2,\theta)\cdot\mathbf{a}(r,\theta)e^{ik_1\Phi(r,\theta)}$, \eqref{eq:12} can be finally rewritten as
\begin{equation}
	\begin{aligned}\label{eq:16}
		\mathbf{e}_2(x,y,z,d)&=-\frac{ik^2_2}{2\pi k^3_1}
		\iint_{r<R}[\mathbf{c}(\phi_1,\phi_2,\theta)e^{ik_0\Psi(\phi_1,\phi_2,d)}e^{ik_{2z}z}/\cos\phi_1]\\
		&\times \exp[-i(k_{1x}x+k_{1y}y)]\,dk_{1x}\,dk_{1y}.
	\end{aligned}
\end{equation}

By using the method developed by Leutenegger \emph{et~al.} \cite{leutenegger2006fast}, we set $|\mathbf{c}|=0$ when $r>R$, the Debye-Wolf intergral is now expressed as the Fourier transform of the field distribution after the interface formed by the two material, ultimately resulting in
\begin{equation}\label{eq:17}
	\mathbf{e}_2(x,y,z,d)=-\frac{ik^2_2}{2\pi k^3_1}\mathcal{F}[\mathbf{c}(\phi_1,\phi_2,\theta)e^{ik_0\Psi(\phi_1,\phi_2,d)}e^{ik_{2z}z}/\cos\phi_1].
\end{equation}

As inspired by Leutenegger \emph{et~al.} \cite{leutenegger2006fast}, we used the chirped Z-transform (CZT) algorithm \cite{Bakx2002} for the Fourier transformation. This algorithm (i) allows breaking the relationship between the sampling points (M) over the aperture radius and the minimal sampling points (N) for fast Fourier transform (FFT), (ii) enables an implicit frequency offset, and (iii) internalizes the zero padding. Applying this generalization, one can adapt the sampling step in the focus field independently of the sampling step in the input field, introduce an additional shift of the region of interest, and finally improve the computational efficiency.

\subsection{Representation of the complex strength vector}
To determine the complex strength vector $\mathbf{c}$, let us assume that the incident field is linearly polarized. By choosing the corresponding Cartesian coordinate and letting the y- and z-components of the incident electric field as zero, the incident electric strength vector can be written as
\begin{equation}\label{eq:18}
	\mathbf{E}^{(0)}=
	\begin{pmatrix}
		E_0\\
		0\\
		0
	\end{pmatrix}.
\end{equation}

According to T\"or\"ok \emph{et~al.} \cite{torok1995electromagnetic}, the transform relation betweeen the incident vector and the refracted vector after the interface can be expressed by a refraction operator $\mathbf{T}$. This operator has a matrix form of 
	\begin{equation}\label{eq:19}
		\mathbf{T}=A(\phi_1)
		\begin{bmatrix}
			a_{11} & a_{12} & a_{13}\\
			a_{12} & a_{22} & a_{23}\\
			-a_{13} & a_{23} & a_{33}
		\end{bmatrix}.
	\end{equation}	
with
\begin{equation*}
	\begin{aligned}
		a_{11} &= \tau_p\cos^2\theta\cos\phi_2+\tau_s\sin^2\theta,\\
		a_{12} &= (\tau_p\cos\phi_2-\tau_s)\cos\theta\sin\theta,\\
		a_{13} &= \tau_p\cos\theta\sin\phi_2,\\
		a_{22} &= \tau_s\cos^2\theta+\tau_p\cos\phi_2\sin^2\theta,\\
		a_{23} &= \tau_p\sin\theta\sin\phi_2,\\
		a_{33} &= \tau_p\cos\phi_2,\\	
	\end{aligned}	
\end{equation*}
where $\tau_s$ and $\tau_p$ are the Fresnel transmission coefficients for s-polarized and p-polarized light, respectively. The function $A(\phi_1)$ is an apodization function that depends on the lens. Moreover when the system obeys Abbe's sine condition, i.e, is aplanatic, then
\begin{equation}\label{eq:20}
	A(\phi_1)=f l_0\sqrt{\cos\phi_1}
\end{equation}
where $f$ is the focal length of the lens in vacuum and $l_0$ is an amplitude factor. It is assumed that the Abbe's sine condition is verified in this step, as it is usually fullfiled in a corrected microscopic lens or in a single spherical lens with the stop located at the lens.

To represent the aberration phase factor and to scale it in wavelength $\lambda$, the aberration factor $e^{ik_1\Phi(r,\theta)}$ is rewritten as $e^{i2\pi W(r,\theta)}$. To describe a circular lens-induced wavefront aberration and calculate the deviation of the wavefront from an ideal spherical shape, one widely used method relies on the Zernike polynomials.  

By using the vendors provided lens data, the aberration function can be calculated in the form of superposition of Zernike polynomials through any homemade ray tracing program or commercial software such as Zemax or Code V. In this paper, we used Zemax \cite{Zemax} to calculate the aberration function and used the notation convention defined by Noll \cite{Noll1976} (so-called Zernike standard polynomials):
\begin{equation}\label{eq:22b}
	W = \sum_{i=1}^{37}c_iZ_i(\rho,\theta_0),
\end{equation}
where $c_i$ is the orthonormal Zernike coefficient computed by Zemax, $Z_i$ is the corresponding Zernike standard polynomial and dimensionless radius $\rho = r/r_{max}$ normalized to the radius of the entrance pupil.

Finally, from \eqref{eq:15}, \eqref{eq:18} and \eqref{eq:22b}, the complex strength vector after the interface can be written as:
	\begin{equation}
		\begin{aligned}\label{eq:23}
			\mathbf{c}&=f l_0\sqrt{\cos\phi_1}\\
			&\times\begin{bmatrix}
				a_{11} & a_{12} & a_{13}\\
				a_{12} & a_{22} & a_{23}\\
				-a_{13} & a_{23} & a_{33}
				\end{bmatrix}
				\begin{bmatrix}
					e^{i2\pi W}&0&0\\
					0&e^{i2\pi W}&0\\
					0&0&e^{i2\pi W}
				\end{bmatrix}
				\begin{bmatrix}
					E_0\\
					0\\
					0
				\end{bmatrix}.
		\end{aligned}
	\end{equation}	

By substituting \eqref{eq:23} into \eqref{eq:12}, a CZT algorithm based highly-efficient propagation model is complete, it allows fast in-focus fields calculation by taking into account both the lens-induced and interface-induced aberrations.

\section{Experimental arrangement}
\subsection{ Propagation imaging}
\begin{figure}
	\centering
	\includegraphics[width=\linewidth]{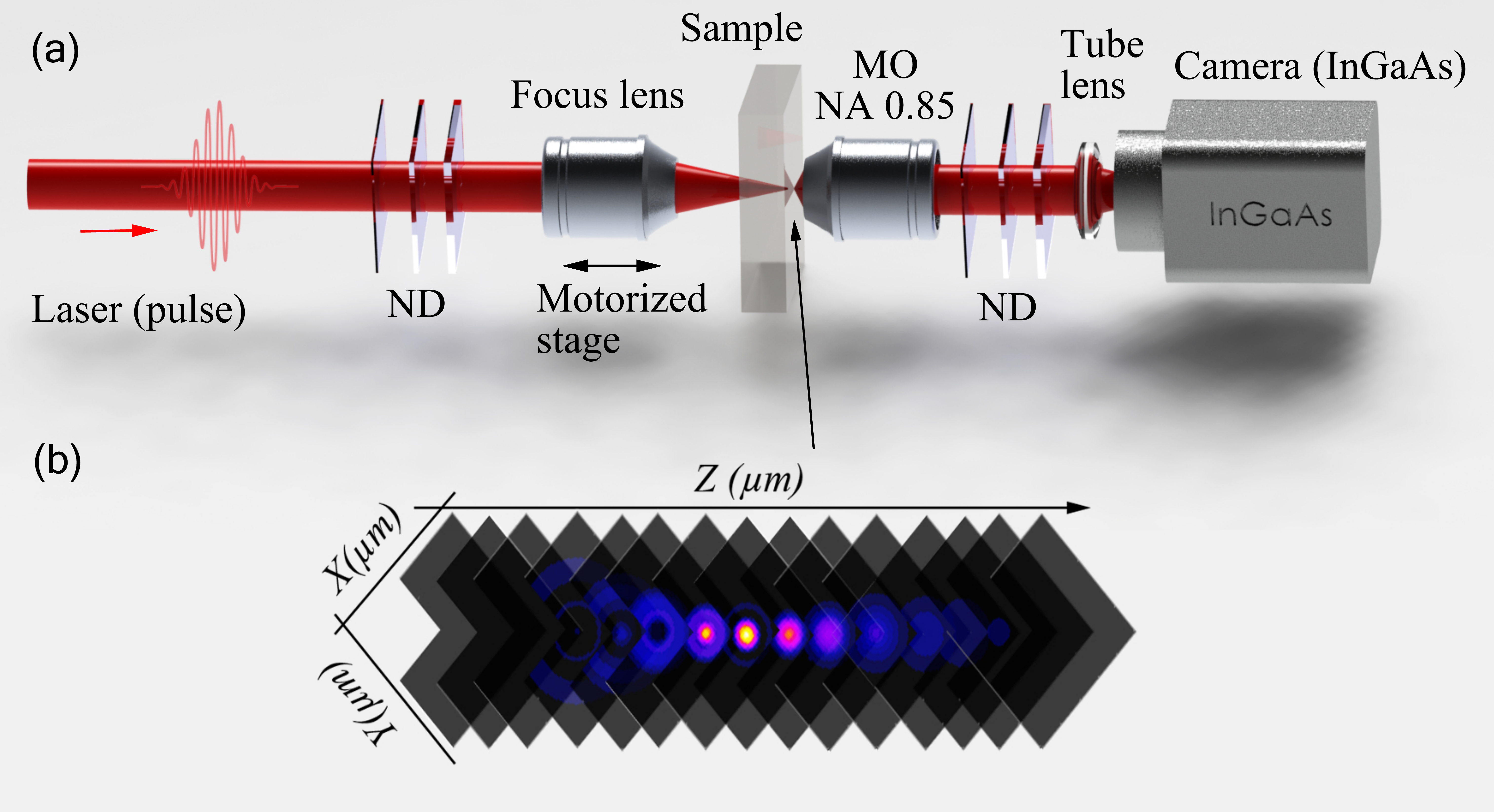}
	\caption{Illustration of in-bulk propagation imaging process. (a) Experimental setup, MO: microscope objective, ND: neutral density filters. (b) Intensity distribution near the focal region reconstructed from the stack of recorded images.}\label{fig:2}
\end{figure}

Propagation imaging methods have been widely used to investigate the linear or nonlinear optical effects in a medium \cite{pasquier2015handling, fedorov2016accessing, wang2020ultrafast,Chambonneau2020}. In this paper, to have a systematic comparison with the numerical results, we introduce non-destructive measurements on the in-bulk volumetric intensity distributions which rely on (i) the focusing of the laser beam at the exit surface of the sample with desired thickness, and (ii) an inverted microscope working in transmission for imaging the beam profile in the \textit{xy} plane for various positions of the focusing objective along the propagation direction \textit{z}. The experimental set-up is schematically depicted in Fig.~\ref{fig:2}.

The measurements can be explicitly divided into two steps. The first step consista in focusing the laser beam with identical characteristics (beam size, spectrum, phase distribution, etc.) as in the simulations. Here, we have chosen a femtosecond pulse laser to avoid the parasitic interference caused by residual multiple reflections. In this ultrashort pulse regime, before focusing, the pulse energy is kept low (typically, a few tens of picojoules) to avoid nonlinear propagation effects. The second step consists in imaging the exit surface of the sample with the inverted microscope working in transmission along the z-axis. This microscope is composed of an infinity-corrected objective lens whose NA is larger than that of the focusing elements, a tube lens and an image sensing array. In this paper, the NA of the imaging objective lens is 0.85 and its focal plane is precisely adjusted at the exit surface of the sample under white light illumination with a translation stage. When there is no sample (focus in air), this plane can be arbitrarily chosen. Thanks to a precision stage (Physik Instrumente, M-126DC1), the laser intensity evolution along the z-axis is recorded by alternating 100-nm movements of the focus lens (corresponding to $n\times100$-nm displacement in a sample with refractive index of $n$), and image acquisitions by the camera. The stack of images is then post-processed for reconstructing the fluence distribution as follows. Firstly, due to the jitter of the laser, some recorded images may show a much higher maximum intensity compare to both the preceding and the following one. These rare outlier images are replaced by the average of the preceding and the following images. Then, in the corner of each image, the noise is calculated as the average of the pixel amplitude, and subsequently subtracted from all pixels. The image containing the maximum gray level is used for the normalization of the whole stack. Two images before and after this latter image are defined for evaluating and correcting the residual tilt of the collecting objective lens with respect to the incoming beam. Finally, the beam propagation is reconstructed by displaying a cross-section (along x or y) of each image at the center of the beam. 

To demonstrate the universality of our method, propagation imaging experiments have been carried out with four different focusing elements and two laser platforms. The first platform is based on an erbium-doped fiber laser (Raydiance Inc, Smart Light 50) with a wavelength $\lambda$ of 1555 nm and a pulse duration of 860 fs. The second one is based on a prototype TRUMPF TruMicro 2030 Femto Edition laser with a wavelength $\lambda$ of 1030 nm and a pulse duration of 265 fs. The average power stability of both lasers is < 1\%. 

\subsection{Materials}
Two singlet lenses (Thorlabs, LA1951-C and C240TME-C) and a 50$\times$ microscope objective lens (Mitutoyo, Apo NIR) are tested with the Raydiance laser platform. The radius of the input Gaussian beam profile is 5.2-mm (at $1/e^2$). A 20$\times$ microscope objective (Mitutoyo, Apo NIR) is tested with the TruMicro 2030 platform, the input beam profiles are shaped through amplitude modulation (slit) or phase modulation (phase plate). To represent the residual aberrations of the two singlet lenses, the nonzero terms of the Zernike standard coefficients are calculated according to our experimental conditions and listed in the Appendix A (Table~\ref{tab:A1}). The laser beams are focused in air ($n$=1) and in crystalline silicon (c-Si, $n$=3.475 at $\lambda$=1555~nm). Detailed information on the materials used in the experiment are listed in Table \ref{tab:1}.
\begin{table*}
	\centering
	\begin{tabular}[c]{l c c c c c c c}
		\toprule
		\rowcolor{black!20}
		Vendor, Lens& f [mm]& NA & $\Phi$ [mm] & $1/e^2$ radius [mm] & $\lambda$ [nm]& Focused-in medium & Presened in\\
		\hline
		Thorlabs, C240TME-C & 8.0 & 0.50 & 8.0 & 5.2 & 1555 & Air & Fig.~\ref{fig:3b}\\
		\rowcolor{black!10}
		Thorlabs, LA1951-C & 25.3 & - & 25.4 & 5.2 & 1555 & Air & Fig.~\ref{fig:3} \\
		Mitutoyo, 20$\times$ Plan Apo NIR & 10.0 & 0.40 & 8.0 & User-defined & 1030 & Air & Fig.~\ref{fig:4}\\
		\rowcolor{black!10}
		Mitutoyo, 50$\times$ Plan Apo NIR HR & 4.0 & 0.65 & 5.2 & 5.2 & 1555 & Si &Fig.~\ref{fig:5}\\
		\bottomrule
	\end{tabular}	
	\caption{Detailed information of the lenses used in this paper. f is the focal length, NA is the numerical aperture, $\Phi$ is the clear aperture diameter, $\lambda$ is the center wavelength of the incident beam.}\label{tab:1}
\end{table*}

\section{Results and discussions} \label{section:3}
In this section we systematically compare our numerical results with the experiments. Under the in-air focus condition, the experimental results benchmarked our numerical model. The further numerical investigations of the in-silicon focus condition, in return, pointed out a limitation of the experimental method.

\subsection{Gaussian beam focused by an aspheric lens in air}
\begin{figure}
	\centering
	\includegraphics[width=\linewidth]{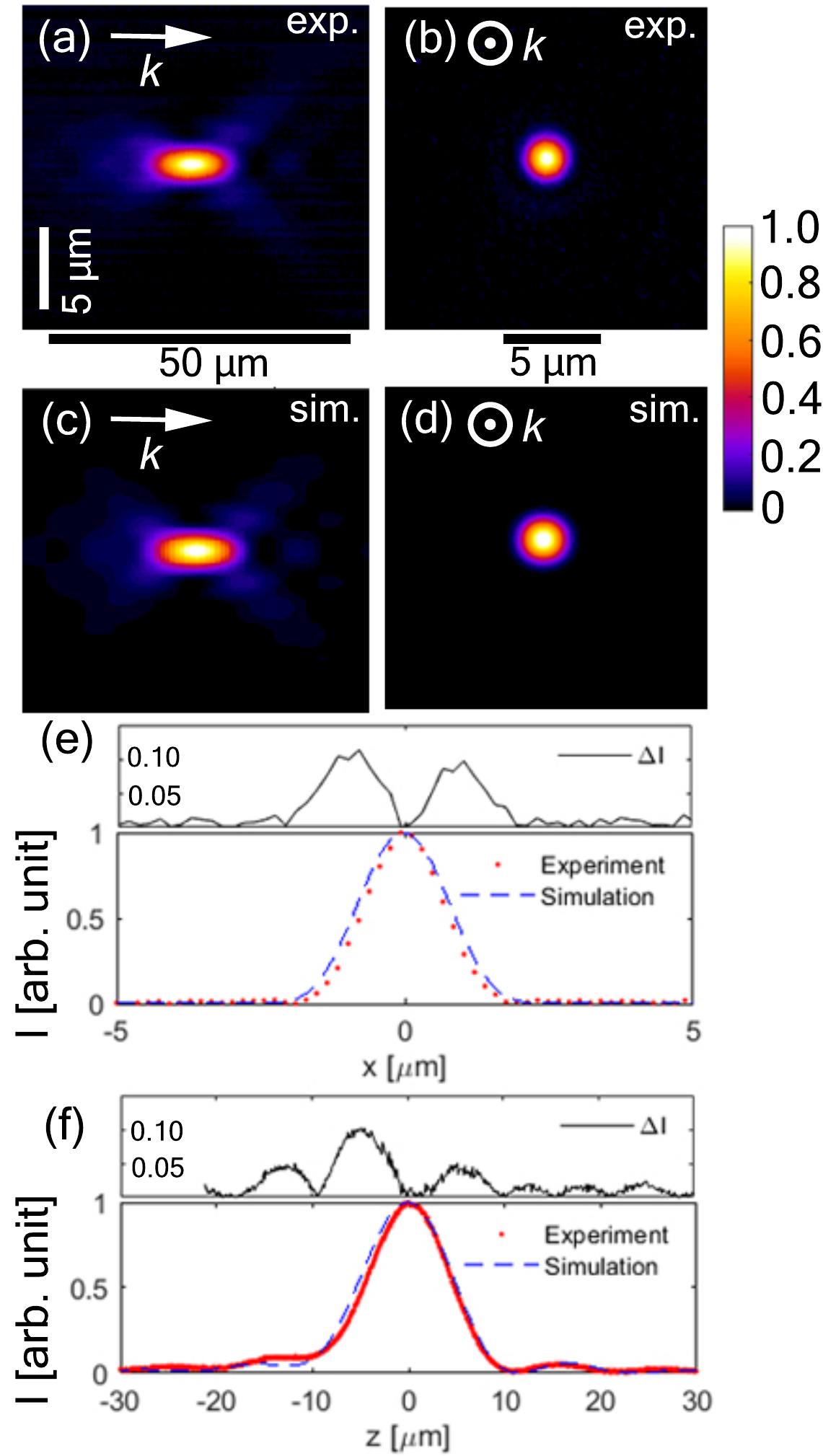}
	\caption{Intensity distribution near the focus of an aspherical lens in air with a Gaussian beam as the input. (a) Experimentally measured intensity along $xz$ plane; (b) experimentally measured intensity at the focal plane; (c) simulated intensity along $xz$ plane; (d) simulated intensity at the focal plane. (e) Comparison of the intensity profiles at the focal plane along the x-axis, (f) along the z-axis.}\label{fig:3b}
\end{figure}

To achieve in-bulk processing, tightly focused laser beams are often required. A cost-efficient solution to get the diffraction-limited high quality tight focusing is to use an aspheric lens. 
In this paper, an aspheric lens (Thorlabs C240TME-C) with NA = 0.5 is used for the first demonstration. As the $1/e^2$ radius of the input beam is 5.2-mm, the lens is overfilled by a factor of 1.3. Therefore, to calculate the aberrations induced by this lens, the apodization factor $G$ (refers to the rate of decrease of the beam amplitude as a function of radial pupil coordinate) is set as $\sqrt{1/1.3}$, i.e., 0.877. The amplitude is normalized to unity at the center of the pupil, and on the other points of the entrance pupil the amplitude is given by $A(\rho)=e^{-G\rho^2}$, where $\rho$ is the normalized pupil coordinate. Under this approximation, the aberrations induced by the lens are calculated and represented by the standard Zernike coefficients. The nonzero coefficients up to the 37th term are listed in Appendix A, Table~\ref{tab:A1}. As a first demonstration, the focused-in medium is air and the corresponding refractive index is 1. Based on \eqref{eq:17} and \eqref{eq:23}, the normalized intensity distribution near the focus is calculated and presented in Fig.~\ref{fig:3b}(c-d). By applying the method presented in section~\ref{section:2}(A), the corresponding experimental results are obtained and presented in Fig.~\ref{fig:3b}(a-b).  
The normalized longitudinal intensity distributions are shown in Fig.~\ref{fig:3b}(a) and (c), and the intensity distributions at the focal plane are shown in Fig.~\ref{fig:3b}(b) and (d). The overall distributions are very similar. To quantitatively evaluate how good the simulation results fit with the experimental one, the normalized experimental intensity profiles are compared to the simulated ones in Fig.~\ref{fig:3b}(e-f), together with the absolute values for the intensity difference. Based on these quantitative analyses, the root-mean-square deviation (RMSD) of those two normalized intensities are calculated for each axis. As shown in Fig.~\ref{fig:3b}(e), by setting the intensity maximized position as the origin, the intensity profiles and their differences are plot from  -5~$\mu$m to 5~$\mu$m. The experimental beam width (at $1/e^2$) is measured as 2.64~$\mu$m, and the simulated one is 2.93~$\mu$m. In this transverse direction, the aforementioned RMSD is calculated as 0.0305. We applied the same methods to the results along the longitudinal direction. The normalized intensity profiles are compared in Fig.~\ref{fig:3b}(f) along the z-axis. The RMSD of the longitudinal profiles is calculated as 0.0322. To anchor a reference, we also simulated the intensity distribution near the focus without considering the influence of the aberrations (not shown here). In this case, the RMSD of the normalized profiles along x, z-axis are 0.0382 and 0.0365. In other words, by taking the aberrations induced by the C240TME-C lens into account, the RMSDs have decreased by 20.2\% and 11.8\% along the x- and z-axis, respectively.

\subsection{Gaussian beam focused by a plano-convex lens in air}
\begin{figure}[t]
	\centering
	\includegraphics[width=\linewidth]{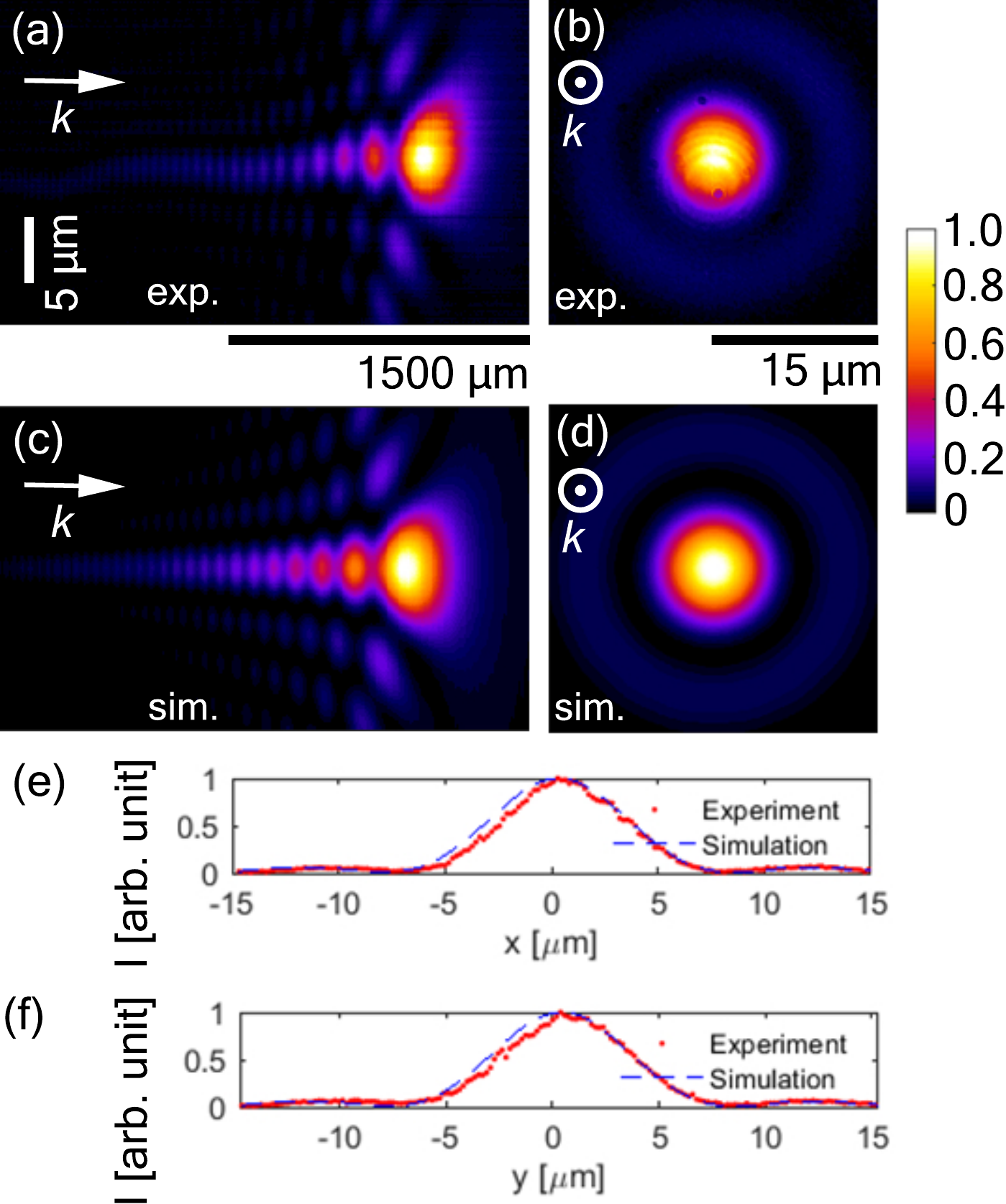}
	\caption{Intensity distribution near the focus of a plano-convex lens in air with a Gaussian beam as the input. (a) Experimentally measured intensity along $xz$ plane; (b) experimentally measured intensity at the focal plane; (c) simulated intensity along $xz$ plane; (d) simulated intensity at the focal plane. Comparisons of the intensity profiles at the focal plane (e) along the x-axis; (f) along the y-axis.}\label{fig:3}
\end{figure}

The plano-convex lens is one of the simplest converging lenses that has been widely used to focus collimated light. The spherical aberrations can be minimized due to the asymmetric design by placing the curved surface face toward the collimated beam, however, it cannot be completely eliminated. In this demonstration, we choose a plano-convex lens (Thorlabs LA1951-C) with $f = 25.3~mm$ and used the two methods mentioned in Section \ref{section:2} and \ref{section:3} to evaluate the intensity distribution at its focus. The same collimated laser beam that was used in the previous section is focused by this lens in air. To calculate the lens-induced aberrations,
the beam amplitude is normalized to unity at the center of the pupil, at other points of the entrance pupil the amplitude is given by $A(\rho)=e^{-G\rho^2}$, where G = 4 and $\rho$ is the normalized pupil coordinate. Under this approximation, the aberrations induced by the lens are calculated and represented by the standard Zernike coefficients. In Appendix A, Table~\ref{tab:A1} the nonzero coefficients are listed up to the 37th term. Finally, the normalized intensity distribution near the focus is calculated based on \eqref{eq:17} and \eqref{eq:23} and presented in Fig.~\ref{fig:3}(c-d). The corresponding experimental results are presented in Fig.~\ref{fig:3}(a-b).

Fig.~\ref{fig:3}(a) and (c) show the normalized longitudinal intensity distributions and Fig.~\ref{fig:3}(b) and (d) shown the intensity distribution at the focal plane. To quantitatively evaluate the agreement, in Fig.~\ref{fig:3} (e-f), we plot the normalized intensity value along the x and y central segment at the focal plane.

As shown in Fig.~\ref{fig:3} (e), along the x-axis, the experimental beam diameter at $1/e^2$ is measured as 10.6~$\mu$m and the simulated one is 11.4~$\mu$m. The RMSD of the two profiles is 0.051. Similar comparisons are applied to the profiles along the y-axis. The experimental beam width is measured as 11.3~$\mu$m, the simulated one is 11.4~$\mu$m and the RMSD is 0.046. From the quantitative comparison one can notice that, in contrast to the simulation results (d), the experimental ones (b) exhibit an elliptical feature. This deviation is due to the fact that for the simulations the incident beam has been oversimplified as a circular Gaussian beam. Apart from that, the simulated results fit well with the experiment. 

It is also worth noting that, given that the experimental results shown in Fig.~\ref{fig:3} (a) are composed by 6000 individual frames, on the left-hand side of this figure some frames are misaligned due to the environmental disturbance such as air flows and vibrations.    
Except for that, one can notice that the aberration features as well as the overall lengths of the focal regime are identical.

\subsection{Exotic beams focused by an objective lens in air}
Microscope objective lenses are also widely used for laser beam focusing and material processing. With a proper design, the aberrations can be well corrected for the design wavelength range. In this section, we choose a 20$\times$ Mitutoyo Plan Apo objective lens (NA = 0.4) as the focus lens. The center wavelength of the laser source is 1030-nm. Instead of using a standard Gaussian beam as the input, we choose amplitude- and phase-shaped beams for the investigation. To precisely simulate the focusing conditions, we first record the intensity profile of the exotic beams at the entrance pupil of the objectives with an array of image sensors, and then calculate the amplitude distribution by taking the square root of the intensity profile. Based on those measured ``user defined'' amplitude profiles and their polarization states, we obtain the complex fields on the entrance pupil. Finally, based on \eqref{eq:17} and \eqref{eq:23}, the fields near the focus can be calculated. These calculated intensity distributions are compared to the experimental results. 

\begin{figure}
	\centering
	\includegraphics[width=\linewidth]{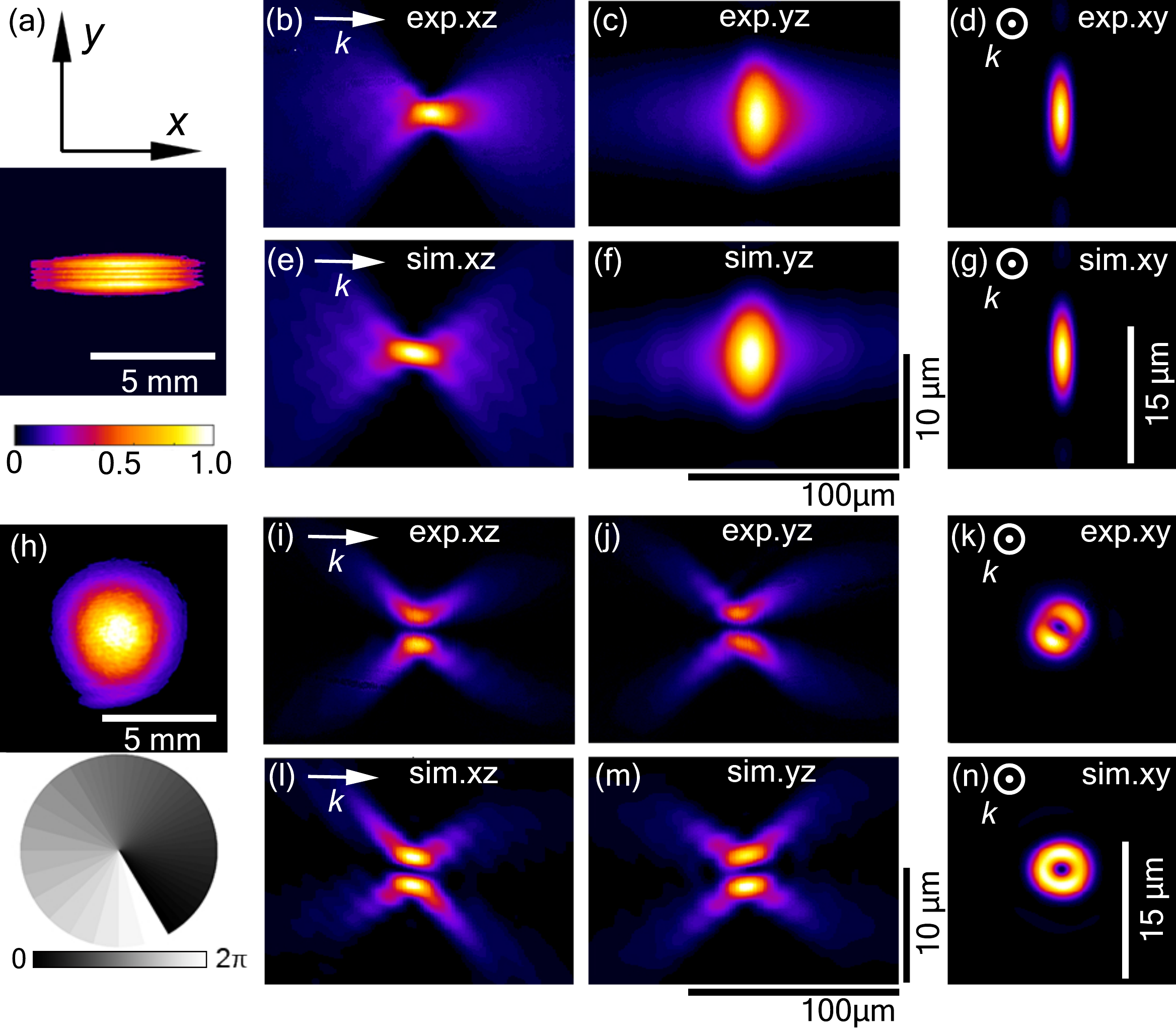}
	\caption{Intensity distribution of the input beam at the entrance pupil for (a) a Gaussian beam cut by a slit with a width of 1.38-mm, (h) a Gaussian beam accumulating a spiral phase (bottom). Intensity distributions near the focus: (b-d) (i-k) experimentally measured along $xz$, $yz$ and $xy$ plane; (e-g) (l-n) simulated along $xz$, $yz$ and $xy$ plane.}\label{fig:4}
\end{figure}

A circular cross-sectional focusing is often required for transversal writing of single-mode waveguides or microfluidic channels. Slit beam shaping is a simple technique that provides such an isotropic resolution in transverse and vertical direction \cite{cheng2003control, ams2005slit}. In this section, we use the slit beam shaping as a first example to prove that our methods are also applicable for the amplitude shaped beam. 

As shown in Fig.~\ref{fig:4} (a), a linearly-polarized Gaussian beam with a $1/e^2$ radius of 5.2~mm is cut by a 1.38-mm width slit. The experimentally measured intensity distributions obtained by focusing this beam are shown in Fig.~\ref{fig:4}(b-d), and the simulated results are shown in Fig.~\ref{fig:4}(e-g). At the focal plane, as shown in Fig.~\ref{fig:4} (d), the beam width along the x-axis is measured as 2.9~$\mu$m and along the y-axis is 10.9~$\mu$m. These beam width values at the focal plane are identical to those obtained in the simulations in Fig.~\ref{fig:4}(g). Similarly, an excellent agreement between the experiments and the calculations is found in the $xy$ and the $yz$ planes, as shown in (b-c) and (e-f), respectively. As expected with such slit beam shaping, the light is focused tighter in the $xz$ plane than in the $yz$ plane. This mainly originates from the loss in the effective numerical aperture along the y-axis due to the fact that the input beam does not overfill the entrance pupil as for the x-axis.
All in all, a near-perfect agreement between the experimental measurements and the calculations demonstrates that our proposed model is a powerful tool even for investigating laser-matter interaction scenarios where sophisticated anisotropic beams are employed. 

However, one should be careful when applying our computational method under extremely asymmetric conditions such as the line-focus microscopy (LTM), as Wolf and Li have indicated that their Debye integral representation should be considered only beyond a critical value of the Fresnel number \cite{wolf1981conditions}. The Fresnel number is a dimensionless number, $\mathcal{N} = a^2/\lambda f$, which reflects the relative contribution of focusing against diffraction effects for a given aperture radius $a$, focal length $f$, and wavelength $\lambda$. For more rigorous approaches dedicated to this specific problem, one can refer to the works of De la Cruz \emph{et~al.} \cite{de2011modeling} and Lou \emph{et~al.} \cite{lou2018better}. They have illustrated that, for Fresnel numbers close to unity, the focus shifts backward, thus leading to astigmatic focusing when the circular symmetry of the input light is broken. In our case where $\mathcal{N}\approx 41$, such a backward shift could not be observed experimentally. 

In the same way as we have studied the influence of amplitude shaping, the impact of phase shaping has also been investigated. Over the last decade, the helical wavefront is one of the most extensively studied complex phase shapes of light. This type of light beams have an azimuthal phase dependence of $\exp(il\theta)$, where $l$ is the topological charge and $\theta$ is the azimuth angle. The optical vortex has many innovative applications in optical tweezers \cite{padgett2011tweezers}, atom manipulation \cite{ladavac2004microoptomechanical} and material processing \cite{hnatovsky2010materials}. When focused, this optical vortex forms a ring instead of a spot in the focal plane. In this section, experimentally, we used a spiral phase plate (SPP) to discretely generate an azimuthal phase distribution of $\exp(-i\theta)$. The topological charge is -1, and the number of discrete steps is 12. After the phase plate, the laser beam is focused by the same 20$\times$ objective lens as previously. In the simulation, to obtain the corresponding complex field as the input, we multiplied the measured amplitude distribution with a discrete spiral phase map that exhibits the same phase distribution as the one used in the experiment. The measured input intensity distribution and the calculated spiral phase map are presented in Fig.~\ref{fig:4} (h).

The simulations in Fig.~\ref{fig:4}(l-n) exhibit similar optical vortex features as the experimental measurements in Fig.~\ref{fig:4}(i-k). From the longitudinal intensity distributions (i-j) and (l-m) one can see that in both cases the light waves along the propagation axis cancel each other out and the ratios of the outer and inner radii are identical. At the focal plane, as shown in Fig.~\ref{fig:4} (k) and (n), this ratio is measured as 4.3 (experimental) and 4.0 (simulated). While, overall, the experimental and theoretical results are in good agreement, one can still note some deviations in the spot elliptically and the intensity homogeneity, most likely provoked by imperfections of the spiral phase plate as well as residual misalignments of this plate.

\subsection{Gaussian beam focused by an objective lens in silicon}
In the previous sections, the laser beams are focused in air. In this section, the in-bulk focus condition is chosen. Unlike in the aforementioned conditions, large experimental to simulation deviation are observed in this scenario. These deviations can be ascribed to the experimental procedure to acquire the intensity profile. 

\begin{figure}[t]
	\centering
	\includegraphics[width=\linewidth]{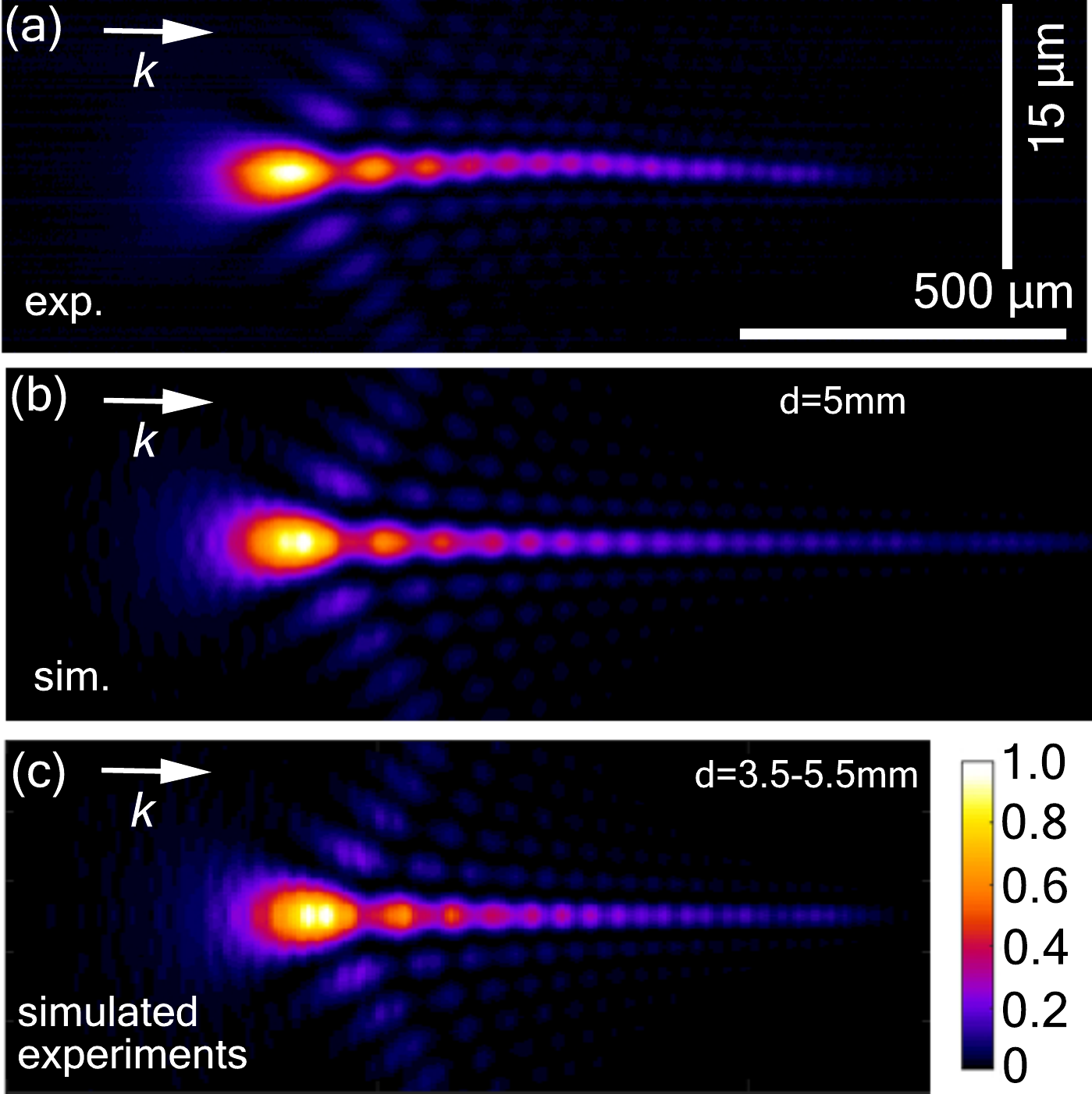}
	\caption{Intensity distributions near the focus, (a) Experimentally measured intensity along $xz$ plane; (b) simulated intensity along $xz$ plane with focal depth of 5-mm; (c) simulated intensity at the exit surface of the 5-mm sample with focal depth varying from 3.5-mm to 5.5-mm.}\label{fig:5}
\end{figure}

When a planar interface is present, according to the aberration function [\eqref{eq:14}] as well as noted in experimental observations \cite{li2016quantitative}, the interface-induced aberration is a function of the focus depth, the NA and the refractive index of the bulk material. By applying the same methodology as described above, we can achieve the experimental and simulated results of the longitudinal intensity distributions near the focus. When a Gaussian beam ($\lambda$: 1555-nm, $1/e^2$ radius: 5.2-mm) is focused into a 5-mm thick crystalline silicon (c-Si) sample ($n$ = 3.475 at 1555-nm) by an objective lens with NA of 0.65, the experimental result is presented in Fig.~\ref{fig:5}(a) and the corresponding simulation is shown in Fig.~\ref{fig:5}(b). Contrary to the previous cases, this comparison shows significant deviations, especially in terms of the overall length. The main reason for this deviation is that, given the employed propagation imaging measurements, the experimental results do not correspond to the intensity distribution at 5-mm focal depth, but it is the stacking of intensity profiles at the exit surface of the 5-mm thick sample for a focusing objective lens moved step-by-step from -1.5 to +0.5 mm (the position 0 corresponding to the exit surface). 

One could wonder why not directly measuring the intensity distribution in the bulk of the material. Unfortunately, it is nearly impossible to measure the real intensity distribution experimentally. This problem cannot be solved by fixing the position of the focusing objective lens and moving the recording microscope since, in this case, the recorded images would be strongly affected by the aberrations provoked by the refractive index mismatch at the back surface of the sample. 
The only way to avoid the artificial result is to compensate the depth related aberration after each movement. Considering 200 movements during one measurement, it is unrealistic to compensate it with any simple means such as a correction collar. However, the experimental result shown in Fig.~\ref{fig:5}(a) can still be exploited for benchmarking our model.

In order to carry out simulations in a situation that is comparable to the experiments, we first calculate the corresponding intensity profile at the exit surface of the 5-mm sample for different focal depths (from 3.5 to 5.5~mm). The focal depth increment step size is 5~$\mu$m. After the calculation, we stack the intensity profiles from the left to the right with the focus depth decrement from 5.5~mm to 3.5~mm. With these two steps, the experimental image acquisition procedure and the corresponding intensity distribution are simulated and displayed in Fig.~\ref{fig:5}(c). While the comparison between the experiments (a) and the simulation for a fixed focus at 5-mm focal depth (b) is mediocre, the experiments and the simulated experiments (c) are in excellent agreement. 

One should emphasize that the simulation at 5-mm focal depth (c) corresponds to the real intensity distribution, which could not be measured experimentally in a simple way. If one aims at measuring the intensity distribution experimentally, extra attention must be paid to the potentially unrealistic character of the results.

\begin{figure}[ht]
	\centering
	\includegraphics[width=\linewidth]{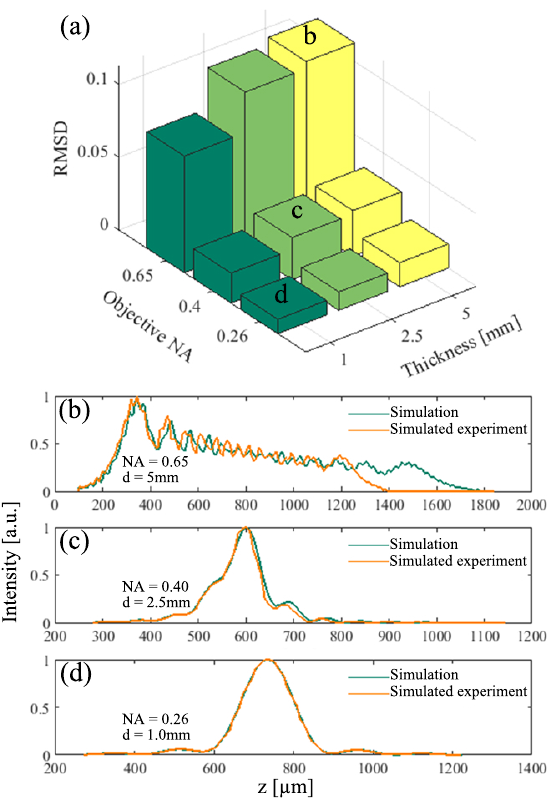}
	\caption{(a) The RMSDs of the simulated actual intensity distribution and the simulated experimental result under different NAs and focal depths. Comparison of the cross-section intensity profiles near the focus along the z-axis with (b) NA = 0.65 and focal depth of 5 mm; (c) NA = 0.4 and focal depth of 2.5 mm; (d) NA = 0.26 and focal depth of 1 mm.}\label{fig:6}
\end{figure}

To quantitatively illustrate the deviation of the experimental measured results and the actual in-bulk intensity distribution, we compare in Fig.~\ref{fig:6} the actual simulated intensity distribution for a fixed focal depth to the simulated experimental results for different numerical apertures and silicon thicknesses.

Given that this deviation is caused by the aberrations induced by the planar interface of the bulk material, the focusing conditions (NA and focal depth) will determine how significant this deviation is. In order to illustrate this relationship and highlight the limitations of propagation imaging experiments, the RMSDs under different numerical apertures and sample thicknesses are calculated in Fig.~\ref{fig:6}(a). As shown in the 3D bar chart, this deviation becomes negligible for small numerical apertures and sample thicknesses. 
In Fig.~\ref{fig:6}(b-d) comparisons of the cross-section profiles along the z-axis are illustrated for three selected cases. When the NA is decreased to 0.26 and the focal depth is decreased to 1 mm, the experimentally measured results are almost identical to the actual in-bulk intensity distribution. The RMSD for this case is 0.009. This value is even smaller than the deviation of the experimental measured and simulated intensity distributions in air, for which there is no influence of the interface-induced aberration. Therefore, in this case, it is safe to use the propagation imaging measurements for representing the actual intensity distribution represent the actual intensity distribution and further apply this propagation imaging method for investigating more complex nonlinear propagation problems \cite{Chambonneau2020}.

\section{Conclusion}
In this paper, we introduced a benchmarked model and an experimental tool for analyzing the field distribution when a laser beam is focused in air or in bulk materials. The vectorial analysis model considered the lens-induced as well as the planar interface-induced aberrations. The in-bulk propagation imaging setup provides a 300-nm longitudinal resolution and diffraction limited lateral resolution. Using the tools introduced in this paper, one can deal with a wide variety of focus conditions in which arbitrary input fields, non-aplanatic lenses, and in-bulk focus might be involved. 

While the numerical simulation tool is applicable, we have also pointed out that for high numerical apertures and significant focal depths, the experimental results acquired by propagation imaging methods that are similar to the one described in this paper might lead to large deviations against the actual intensity distribution in the desired focal depth. One should calculate the RMSD or at least check the RMSD chart before applying these methods to any linear or nonlinear propagation imaging experiments. 

We plan to continue the development of InFocus \cite{InFocus} in the spirit of open-source and would be pleased to find collaborators. We anticipate that our proposed model and corresponding software InFocus can be utilized in countless laser processing applications involving various wavelengths, beam shapes, phase distributions and focusing optics.


\appendix
\section*{Appendix A: Zernike standard coefficients of the single lens}
\setcounter{table}{0}
\renewcommand{\thetable}{A\arabic{table}}
This appendix lists the nonzero terms of the Zernike standard coefficients calculated at the exit pupil of the single lens LA1951-C and C240TME-C. For both lenses, the properties of the incident laser beams are the same, i.e., a wavelength of 1555 nm and a $1/e^2$ radius of 5.2 mm. The refractive index of the medium after the lens is 1.
\begin{table}[h!]
	\centering
	\begin{tabular}[c]{l c c c }
		\toprule
		\rowcolor{black!20}
		 & LA1951-C & C240TME-C & Polynomials\\
		$Z_1$ & 2.484 & 0.5821 & 1 \\
		\rowcolor{black!10}
		$Z_4$ & 2.171 & 0.3187 & $\sqrt{3}(2r^2-1)$ \\
		$Z_{11}$ & 0.581 & -0.0209 & $\sqrt{5}(6r^4-6r^2+1)$ \\
		\rowcolor{black!10}
		$Z_{22}$ & 0.009 & -0.007 & \vtop{\hbox{\strut $\sqrt{7}(20r^6-30r^4$}\hbox{\strut $+12r^2-1)$}}\\
		$z_{37}$ & 0.0002 & -0.007 & \vtop{\hbox{\strut $\sqrt{9}(70r^8-140r^6$}\hbox{\strut $+90r^4-20r^2+1)$}}\\
		\bottomrule
	\end{tabular}	
	\caption{Nonzero terms of the Zernike standard coefficients calculated at the exit pupil of a plano-convex lens Thorlabs LA1951-C and an aspherical lens C240TME-C.}\label{tab:A1}
\end{table}

\begin{backmatter}
	\bmsection{Acknowledgment} This research has been supported by the Bundesministerium für Bildung und Forschung (BMBF) through the NUCLEUS project, grant no. 03IHS107A, as well as the glass2met project, grant no. 13N15290.
	\bmsection{Disclosures} The authors declare no conflicts of interest.
\end{backmatter}


\bibliography{references}

\end{document}